\documentclass[letterpaper]{jpconf}
\usepackage{graphicx}
\usepackage{epsf}
\usepackage{amsfonts}
\usepackage{amsmath}
\usepackage[mathscr]{eucal}
\usepackage{booktabs}

\newcommand{\cP}{{\cal P}}                  

\begin{document}

\title{Chiral polarization scale of QCD vacuum and spontaneous chiral symmetry breaking}

\author{Andrei Alexandru$^{1}$ and Ivan Horv\'ath$^{2,}$\footnote[3]{The Speaker.}}

\address{$^{1}$George Washington University, Washington, DC, USA}

\address{$^{2}$University of Kentucky, Lexington, KY, USA}

\ead{aalexan@gwu.edu, horvath@pa.uky.edu}

\begin{abstract}
It has recently been found that dynamics of pure glue QCD supports the low energy
band of Dirac modes with local chiral properties qualitatively different from that 
of a bulk: while bulk modes suppress chirality relative to statistical independence
between left and right, the band modes enhance it. The width of such chirally polarized 
zone -- {\em chiral polarization scale} $\Lambda_{ch}$ -- has been shown to be finite 
in the continuum limit at fixed physical volume. Here we present evidence that 
$\Lambda_{ch}$ remains non--zero also in the infinite volume, and is therefore a dynamical 
scale in the theory. Our experiments in N$_f$=2+1 QCD support the proposition
that the same holds in the massless limit, connecting $\Lambda_{ch}$ 
to spontaneous chiral symmetry breaking. In addition, our results suggest
that thermal agitation in quenched QCD destroys both chiral polarization and condensation 
of Dirac modes at the same temperature $T_{ch} > T_c$.
\end{abstract}

\vskip -0.2in
\section{Introduction}
The work presented in this talk is part of a long--term effort to look at QCD vacuum
structure in a model--independent manner. This involves devising meaningful, preferably
gauge invariant, characteristics of gauge configurations dominating QCD path integral.
These configurations are accessible via first--principles lattice QCD simulations, and
are not altered from their equilibrium form. The underlying assumption is that when 
objective information of this kind sufficiently accumulates, it can be
eventually integrated into a coherent picture of vacuum--related QCD phenomena. This 
is the {\em bottom--up approach} to QCD vacuum structure~\cite{Hor06A}. 

Our focus here is spontaneous chiral symmetry breaking (SChSB), generally viewed 
as one of the keys to making sense of low--energy hadronic physics. When thinking
about SChSB, Dirac eigenmodes immediately pop out. Indeed, spectral representation 
turns out to be a fruitful way of looking at $\bar{\psi}\psi$, revealing that symmetry 
breakdown is equivalent to Dirac mode condensation. For example, in N$_f$=2 theory 
with quark mass $m$,  
\begin{equation}
\lim_{m\to 0} \langle \bar{\psi} \psi \rangle_m \ne 0 
\quad \Longleftrightarrow \quad
\lim_{\lambda\to 0}\lim_{m\to 0} \rho(\lambda,m) \ne 0
\label{eq:10}
\end{equation}
where $\rho(\lambda,m)$ is the spectral density of modes and we assumed infinite 
volume~\cite{Ban80A}. While the above clarifies meaning of chiral symmetry in spectral 
language, it offers no dynamical detail on the breaking phenomenon. The underlying 
goal of this project is to identify such detail in local behavior of Dirac modes.

Our initial interest in this direction are {\em chiral properties} of the modes,
owing to the intuitive expectation that breakdown of chiral symmetry should be 
imprinted in such features. While global chirality of Dirac non--zeromodes vanishes, 
the local chiral behavior reflects properties of the underlying gauge 
background~\cite{Hor01A}. The simplest ``local inquiry'' of the above type is 
whether values $\psi(x)=\psi_L(x) + \psi_R(x)$ tend to appear with equal participation 
of left/right subspaces or with asymmetric one. In other words, whether they tend 
to be chirally polarized or anti--polarized. Such questions were asked some time ago 
with different goals in mind~\cite{Hor01A}, but it turns out that for our purposes 
it is crucial that the corresponding measures be {\em dynamical}.  
In the current context, ``dynamical measure'' is one that it is uniquely defined
and involves comparison to statistical independence of left and right. 
Such measures have been constructed in Ref.~\cite{Ale10A}. 

\begin{figure}[t]
\begin{center}
    \centerline{
    \hskip 0.14in
    \includegraphics[width=6.0truecm,angle=0]{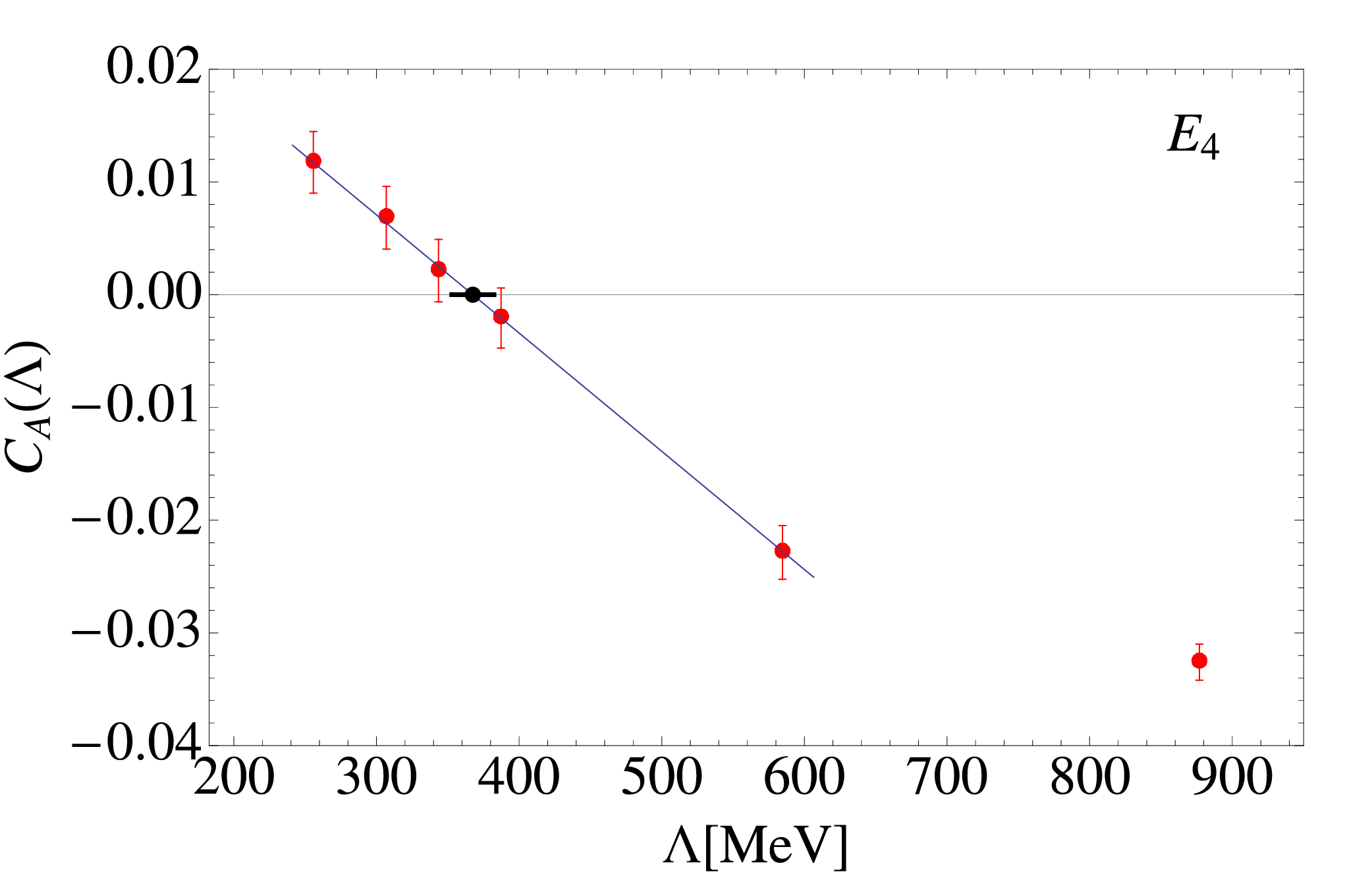}
    \hskip -0.11in
    \includegraphics[width=5.6truecm,angle=0]{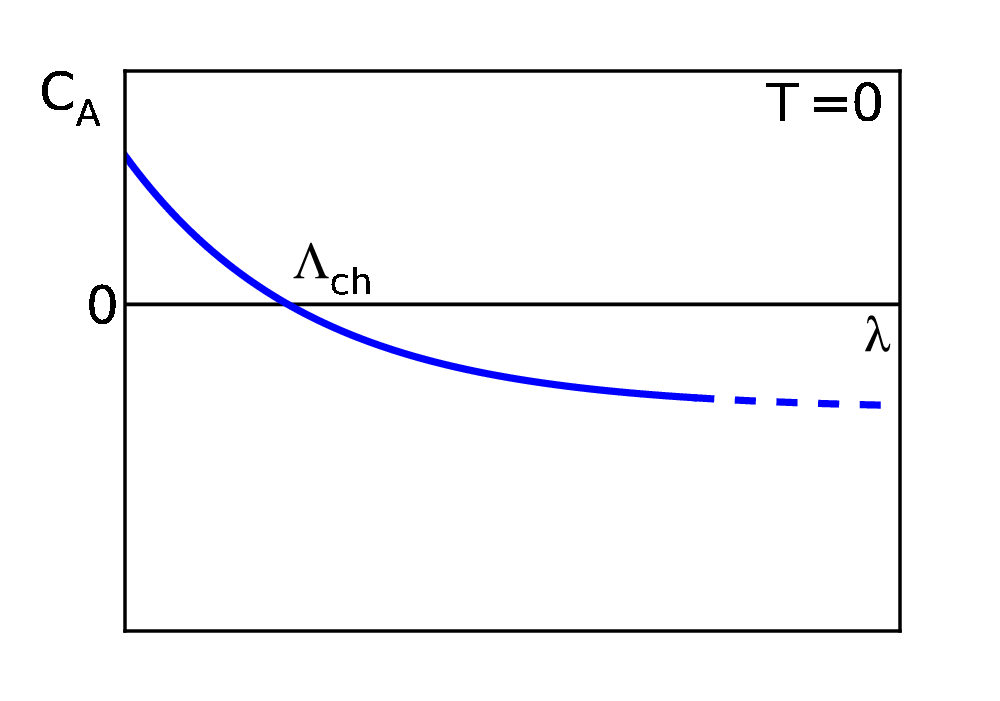}
    \hskip -0.23in
    \includegraphics[width=5.6truecm,angle=0]{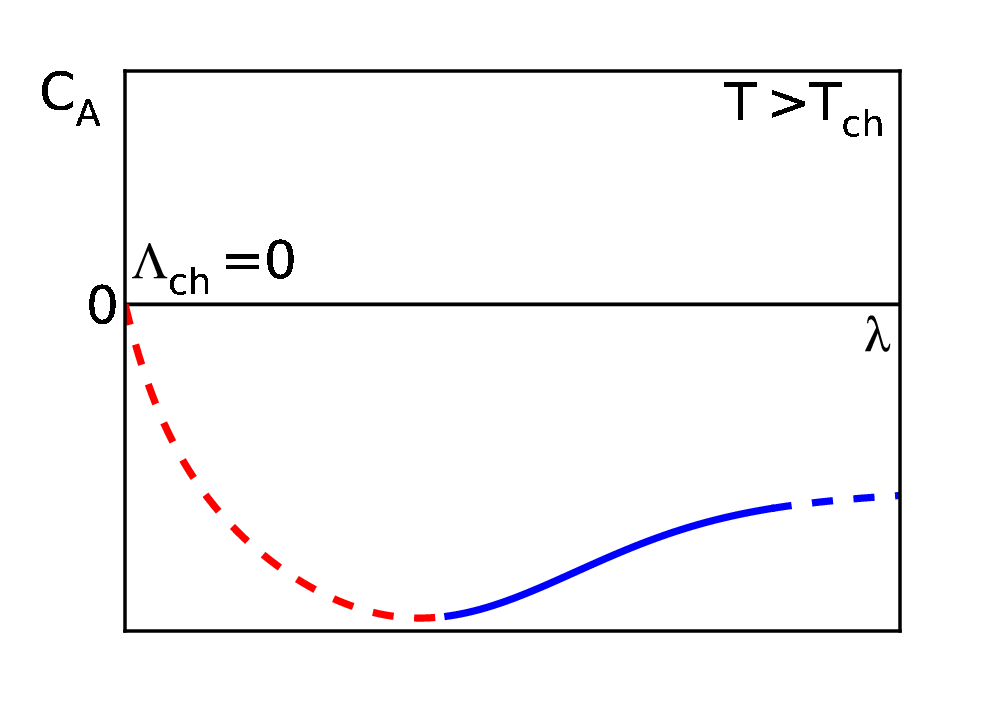}
     }
     \vskip 0.00in
     \caption{Left: chiral polarization scale as found in ensemble E$_4$ of Ref.~\cite{Ale10A}.
     Middle: qualitative behavior of $C_A(\lambda)$ in zero--temperature QCD. Right: same after
     chiral symmetry restoration by thermal agitation.}
    \label{fig:illus}
    \vskip -0.50in
\end{center}
\end{figure}

In this work we will only deal with the most basic polarization measure, namely 
the correlation coefficient of chiral polarization $C_A$. This quantifies
the overall tendency toward local chiral asymmetry, and is constructed 
as follows. Consider the probability distribution $\cP (\psi_L,\psi_R)$ of 
left--right components for a mode or some group of modes. Associated with $\cP$
is the distribution $\cP^u (\psi_L,\psi_R) \equiv P(\psi_L) P(\psi_R)$ 
describing statistically independent components. 
Here $P(\psi_L) = \int d \psi_R \cP(\psi_L,\psi_R)$ and similarly for $P(\psi_R)$,
with functional forms being the same due to the symmetry of $\cP$. Now, imagine 
the experiment in which samples are being drawn simultaneously from $\cP$ and
$\cP_u$, and compared by their polarization. The outcomes of such comparisons have
no arbitrariness, and their statistics defines the probability $\Gamma_A$ that 
sample chosen from $\cP$ is more polarized than one chosen from $\cP^u$. 
The correlation coefficient is then $C_A \equiv 2 \Gamma_A -1 \in [-1,1]$. 
Consequently, the dynamics enhancing polarization relative to statistical 
independence is chirally correlated ($C_A>0$) while the one suppressing it is 
anti--correlated ($C_A < 0$): the former supports local chirality while the latter
local anti--chirality.

The new dynamical information associated with the above approach is encoded in spectral 
behavior of $C_A \equiv C_A(\lambda)$, with spectral average in finite volume
defined by~\cite{Ale12A} 
\begin{equation}
   C_A(\lambda,M,V) \equiv 
   \frac{\sum\limits_k \langle \, \delta(\lambda - \lambda_k) \, C_{A,k} \,\rangle_{M,V}}
        {\sum\limits_k \langle \, \delta(\lambda - \lambda_k) \, \rangle_{M,V}}
   \label{eq:20}
\end{equation}
Here $M\equiv (m_1,m_2,\ldots,m_{N_f})$ is the set of quark masses and $C_{A,k}$ 
the correlation associated with k-th mode. In Ref.~\cite{Ale10A} it was found that 
$C_A(\lambda)$ in quenched QCD has a positive core around zero, and switches to negative 
values at chiral polarization scale $\Lambda_{ch}$ (Fig.\ref{fig:illus}, left). It was 
also shown that $\Lambda_{ch}$ is non--zero in the continuum limit at fixed physical 
volume. Here we present evidence that (1) $\Lambda_{ch}$ remains positive in 
the infinite volume limit and is thus a dynamical scale in the theory, (2) $\Lambda_{ch}$
is non--zero in the chiral limit of $N_f=2+1$ QCD and SChSB thus proceeds via chirally 
polarized modes, and (3) $\Lambda_{ch}$ vanishes simultaneously with density of 
near--zeromodes when temperature is turned on, and is thus a scale closely tied to SChSB. 
Qualitative behavior of $C_A(\lambda)$ in zero--temperature QCD is shown in the middle
plot of Fig.\ref{fig:illus}, while that in the chirally symmetric phase at finite
temperature is on the right.

\begin{figure}[t]
\begin{center}
    \centerline{
    \hskip 0.05in
    \includegraphics[width=7.8truecm,angle=0]{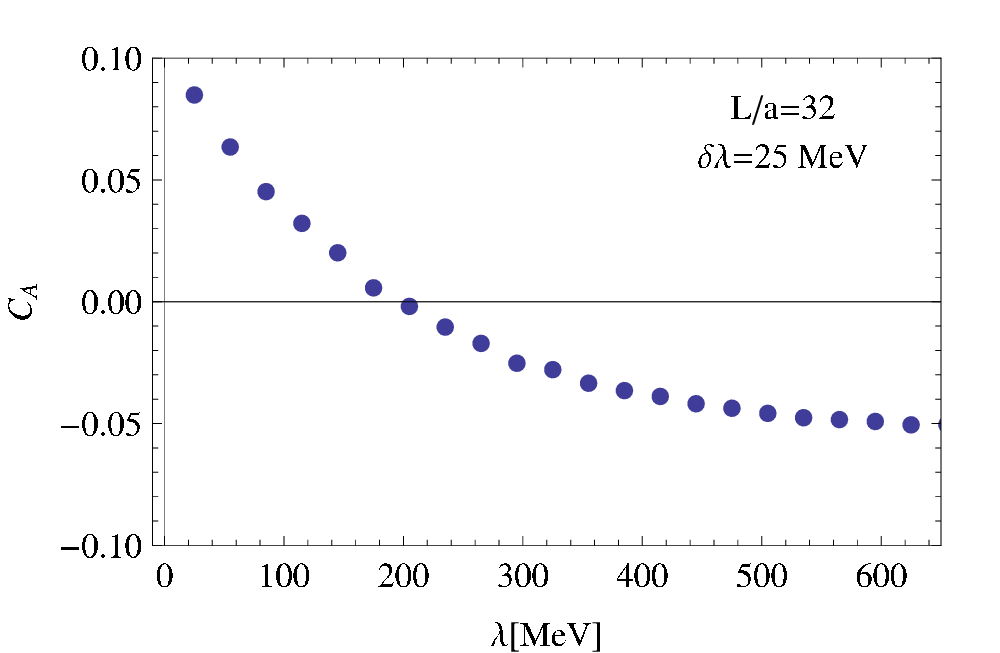}
    \hskip -0.05in
    \includegraphics[width=7.8truecm,angle=0]{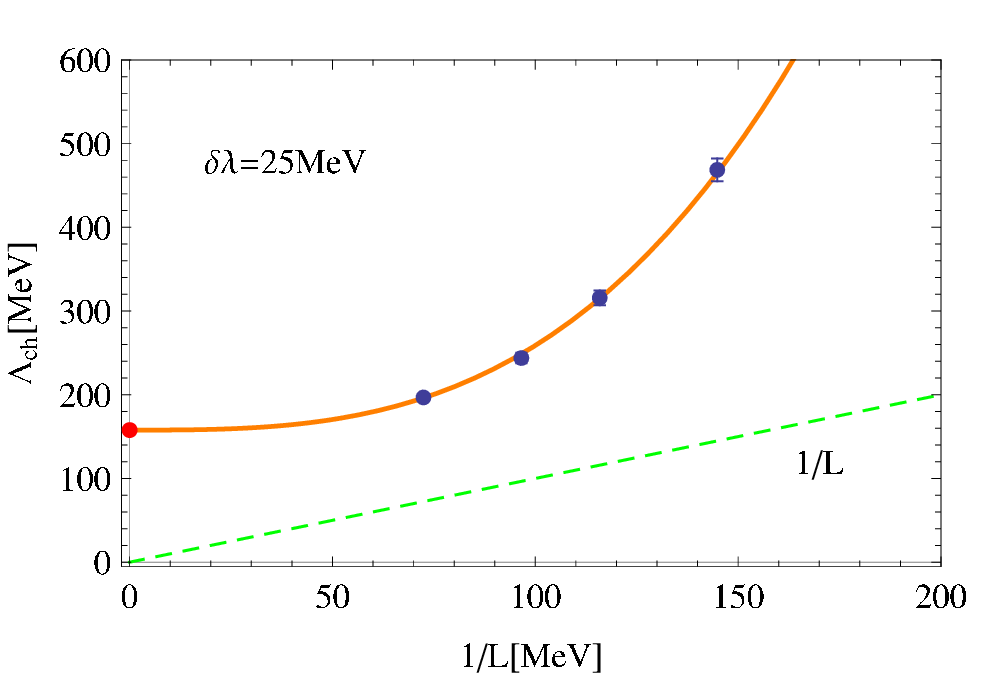}
     }
    \vskip -0.00in
    \centerline{
    \hskip 0.05in
    \includegraphics[width=7.8truecm,angle=0]{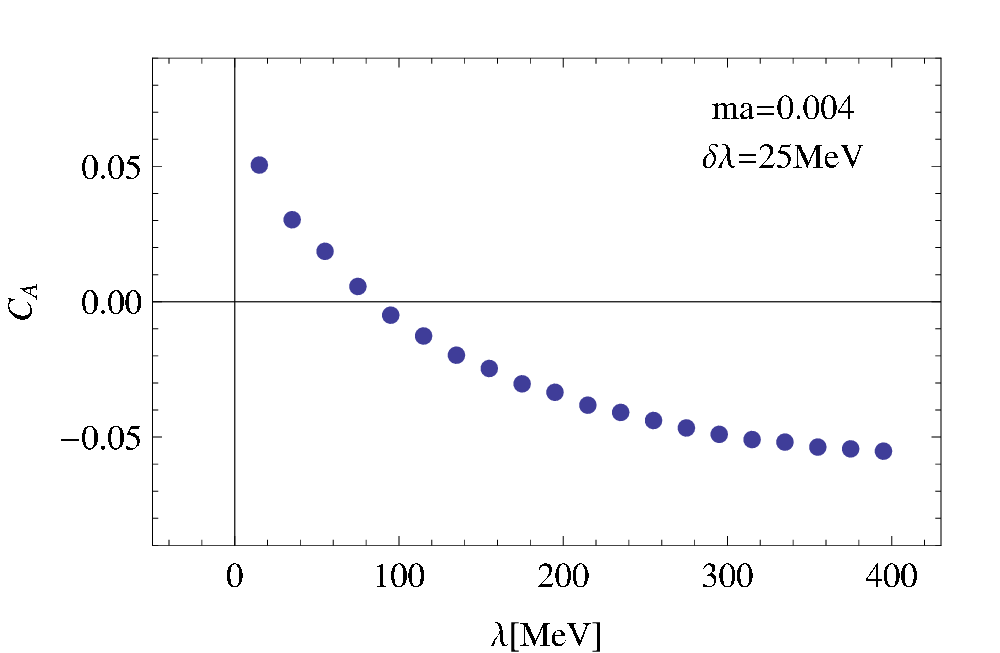}
    \hskip -0.05in
    \includegraphics[width=7.8truecm,angle=0]{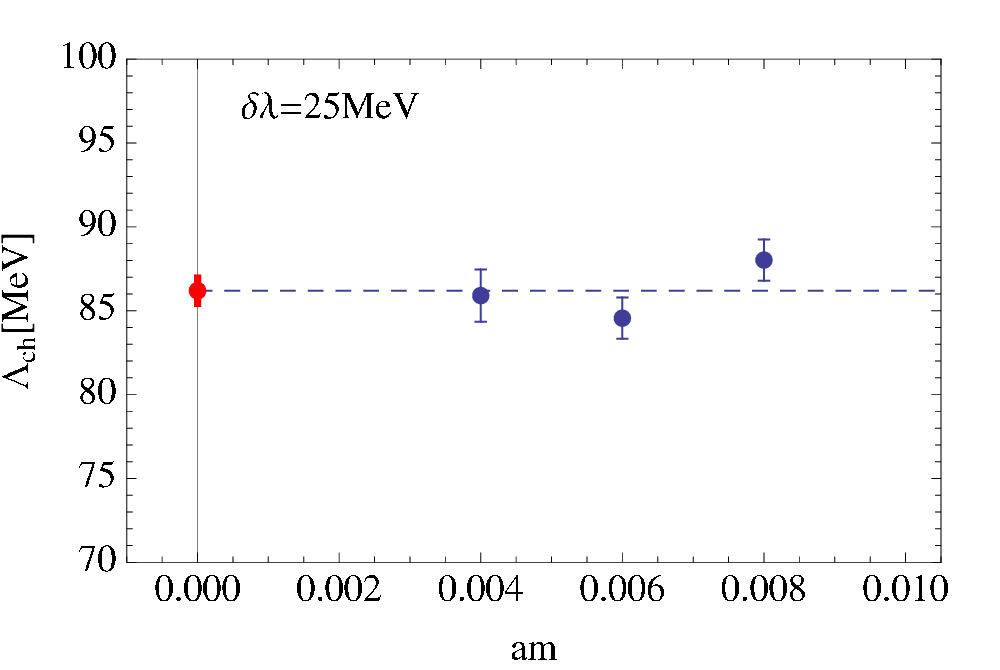}
     }
     \vskip 0.00in
     \caption{Top: $C_A(\lambda)$ for $L=32$ lattice of quenched QCD (left) and 
     infinite volume extrapolation (right). Bottom: $C_A(\lambda)$ for $m_l$=0.004
     lattice of $N_f$=2+1 QCD (left) and chiral limit extrapolation (right).} 
    \label{fig:inf_chir}
    \vskip -0.45in
\end{center}
\end{figure}

\section{Infinite volume and chiral limit.}
We start with the question of infinite volume limit. As pointed out in Ref.~\cite{Ale10A}, 
this is of crucial importance in order to establish that $\Lambda_{ch}$ represents 
dynamically generated scale of QCD. We perform the volume study in the context of pure glue
lattice QCD with Wilson action at fixed gauge coupling $\beta$=6.054. Using 
the parametrization of Ref.~\cite{Gua98A} and value $r_0$=0.5 fm this translates into 
the lattice spacing $a$=0.085 fm. Lattices of sizes 16$^4$, 20$^4$, 24$^4$ and 32$^4$
were studied with 100 independent configurations generated in each case.

To probe local chiral properties of modes, we use overlap Dirac operator with parameters 
$r$=1 and $\rho$=26/19, and these settings were used throughout this study. Approximately 200 
conjugate pairs of lowest near--zeromodes were computed on each configuration, and 
the average $C_A(\lambda)$ was determined in the available region of the spectrum. 
In Fig.\ref{fig:inf_chir} (top left) we show the result for the largest volume, exhibiting 
the anticipated behavior, and with clearly defined chiral polarization scale. Dependence 
of $\Lambda_{ch}$ on the infrared cutoff is shown in the top right plot of 
Fig.\ref{fig:inf_chir}, together with the cutoff $1/L$ itself. The data clearly curves away 
from the cutoff, indicating a positive infinite volume limit. The fit of the form 
$\Lambda_{ch}(1/L) = \Lambda_{ch}(0) + b\, (1/L)^3$
was used to extrapolate to infinite volume, yielding $\Lambda_{ch} \approx 160$ MeV at this 
cutoff. Judging by the behavior of continuum extrapolation in Ref.~\cite{Ale10A},
the continuum limit is estimated to be $\Lambda_{ch} \approx 150$ MeV in Wilson 
regularization of quenched QCD.

To study the effects of dynamical fermions, we use the N$_f$=2+1 {\bf RBC/UKQCD} domain wall 
fermion ensembles of Ref.~\cite{RBC09A}. These 32$^3 \times$64 lattices are at fixed heavy 
bare mass $m_h a\!=\!0.03$, approximately corresponding to mass of physical strange quark,
and at fixed lattice cutoff $a\!=\!0.085$ fm. 
The light quark masses $m_l a\!=\!0.004,0.006,0.008$ reach the lightest pion mass 
of $295$ MeV. Low--lying overlap Dirac eigenmodes were computed on 50 configurations 
from each ensemble, with functions $C_A(\lambda)$ evaluated on the spectral 
region available. The result for lightest quark mass is displayed on the bottom left plot of 
Fig.~\ref{fig:inf_chir}, showing behavior qualitatively similar to quenched case.
Chiral polarization scale has been reduced due to the effects of light dynamical quarks,
but the sensitivity to light mass has already become negligible in this region, as seen on 
the bottom right plot of Fig.~\ref{fig:inf_chir}. Positive chiral limit, indicated via 
extrapolation by a constant, is thus expected. The preferred value is 
$\Lambda_{ch} \approx 86$ MeV at this cutoff, with rough estimate of 
$\Lambda_{ch} \approx 80$ MeV in the continuum. Note that, due to flat mass dependence,
the above estimates apply both at physical point and in the chiral limit.  

\begin{figure}[t]
\begin{center}
    \centerline{
    \hskip 0.05in
    \includegraphics[width=7.0truecm,angle=0]{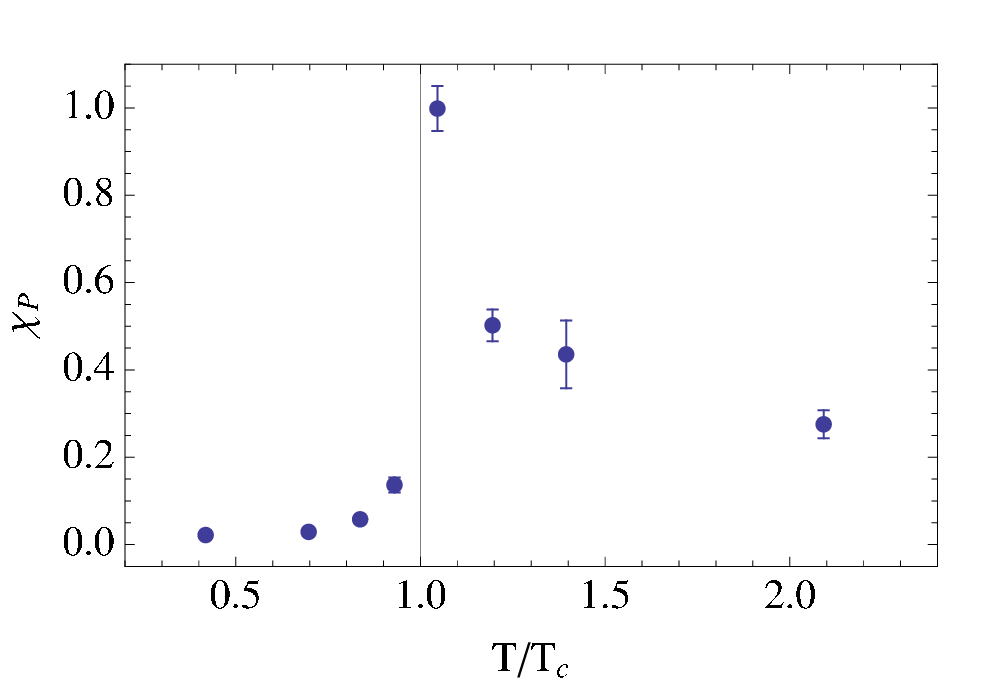}
    \hskip -0.00in
    \includegraphics[width=7.0truecm,angle=0]{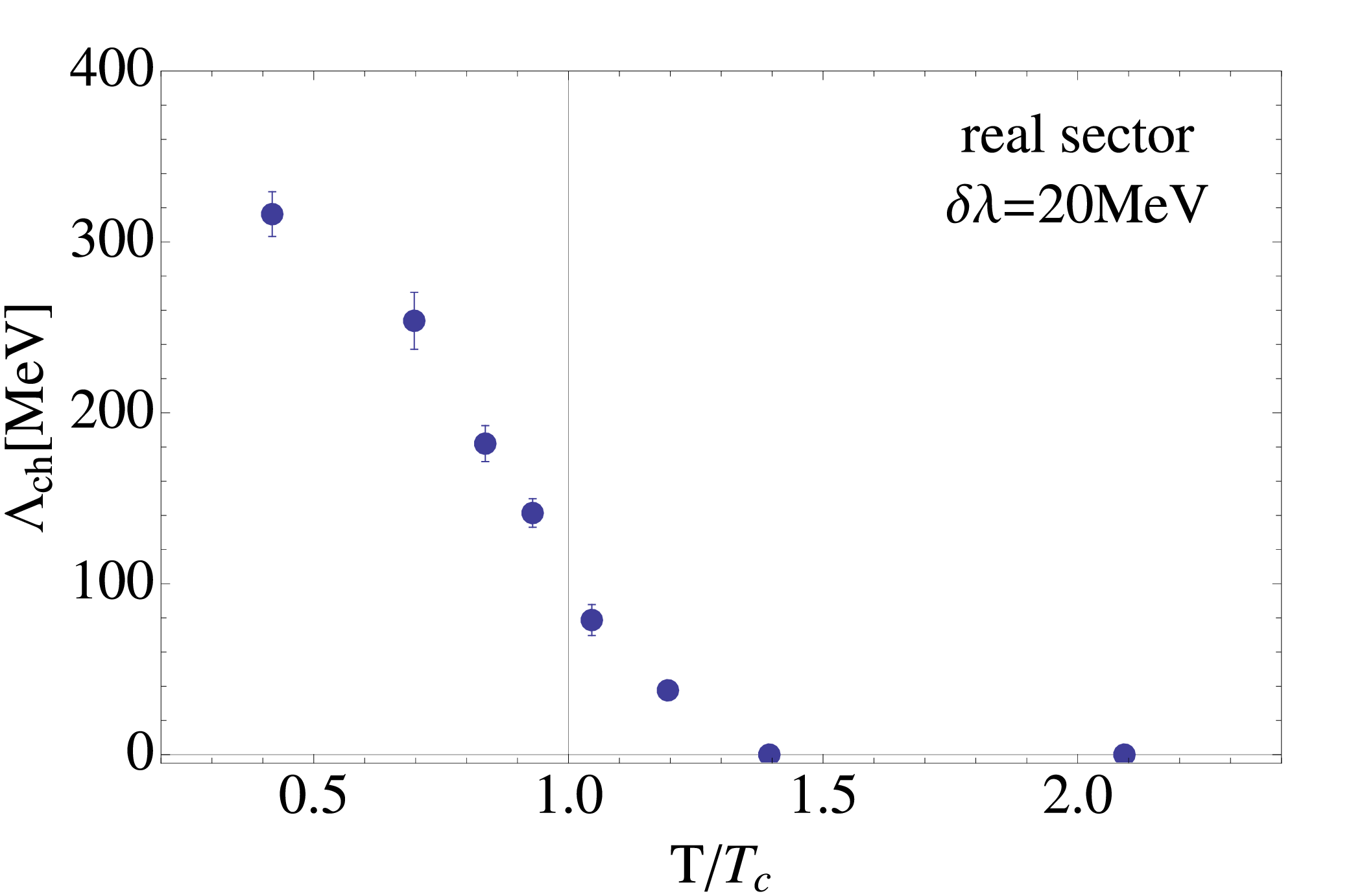}
     }
    \vskip -0.02in
    \centerline{
    \hskip 0.05in
    \includegraphics[width=7.0truecm,angle=0]{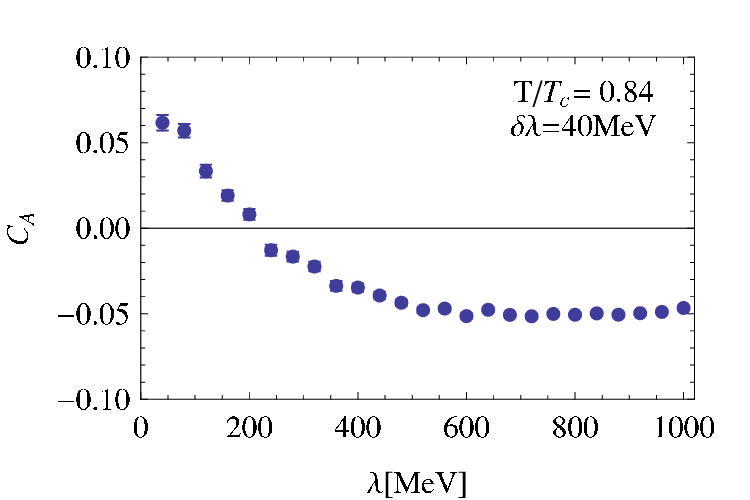}
    \hskip -0.00in
    \includegraphics[width=7.0truecm,angle=0]{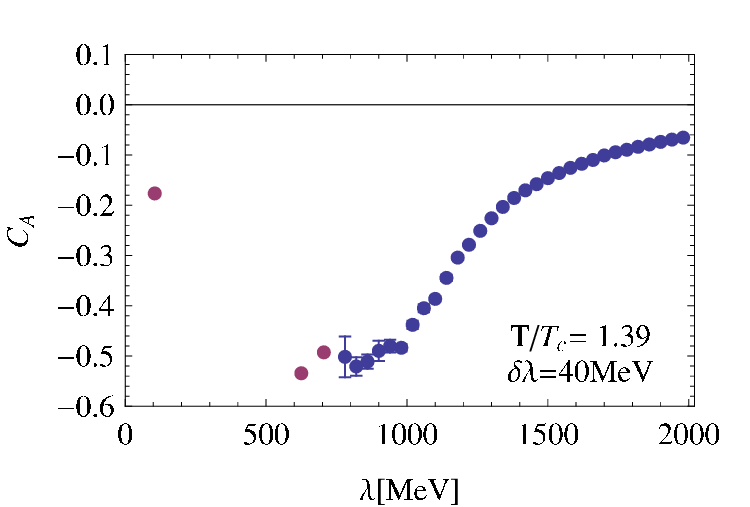}
     }
     \vskip 0.00in
     \caption{Top: temperature dependence of Polyakov line susceptibility (left)
     and chiral polarization scale (right) as discussed in the text. Bottom: behavior 
     of $C_A(\lambda)$ at $T/T_c=0.84$ (left), and at $T/T_c=1.39$ (right).}
    \label{fig:fin_temp}
    \vskip -0.45in
\end{center}
\end{figure}

\section{Chiral transition} 

We now wish to examine whether low--energy chiral polarization could be tied to 
chiral symmetry breaking in yet more fundamental way. In particular, we study whether 
the condensation of Dirac modes ($\rho(\lambda \to 0)>0$) and chiral polarization 
($\Lambda_{ch}>0$) tend to occur simultaneously~\cite{Ale12A}. To that end, we 
consider quenched QCD at finite temperature $T$ where it is expected that, 
for $T>T_{ch}$, Dirac modes cease to condense, and ``valence'' chiral symmetry becomes 
restored. Note that the chiral transition temperature was denoted as $T_{ch}$, 
in order to distinguish it from the deconfinement temperature $T_c$ defined by 
the breakdown of $Z_3$ symmetry.

We use Wilson lattice regularization again, at the identical gauge coupling 
$\beta$=6.054. Overlap Dirac eigenmodes were computed on $20^3 \times N_t$ lattices 
with $N_t=20,12,10,9,8,7,6,4$, spanning both confined and deconfined regions. 
In the top left panel of Fig.~\ref{fig:fin_temp} we plot the temperature dependence 
of Polyakov line susceptibility, showing expected behavior with peak at 
$T/T_c=1.05$ ($N_t$=8). The value of $T_c$, including all extrapolations, is taken 
from Ref.~\cite{Kar97A}. Note that for spectral calculations, only configurations 
from the ``real $Z_3$'' Polyakov line sector were used~\cite{Cha95A}, so that 
the continuation to theory with dynamical quarks is smooth.

\begin{figure}[t]
\centerline{
\includegraphics[width=15pc]{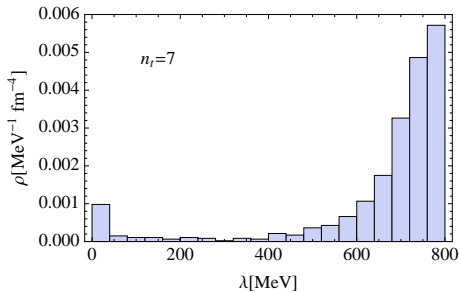}\hspace{2pc}%
\begin{minipage}[b]{17pc}\caption{\label{label}Detail of mode density at low end of the
spectrum for $N_t$=7. Isolated band of near--zero modes is clearly visible. Zero modes
are not included.}
\label{fig:remnant}
\end{minipage}
}
\vskip -0.12in
\end{figure}

In Fig.~\ref{fig:fin_temp} (bottom) we show the computed $C_A(\lambda)$ at 
$T/T_c\!=\!0.84$ ($N_t\!=\!10$), and at $T/T_c\!=\!1.39$ ($N_t=6$). In the former 
(confined) case there is a positive core of chiral polarization, while only 
anti--polarization exists in the (deconfined) latter.\footnote{The red points 
in the plot for $T/T_c=1.39$ denote bins with not enough eigenmodes for error 
to be estimated.} Closer inspection reveals a discrepancy of chiral polarization 
behavior with respect  to confinement, but not with respect to mode condensation. 
Indeed, as can be seen
in Fig.~\ref{fig:fin_temp} (top right), chiral polarization is present 
($\Lambda_{ch}>0$) at $T/T_c$=1.2 ($N_t=7$), i.e. quite safely in the deconfined phase. 
At the same time though, as the profile of mode density in Fig.~\ref{fig:remnant}
suggests, $\rho(\lambda)>0$ in the narrow band close to zero, consistently with
condensation and breakdown of valence chiral symmetry. The residual density of this 
kind was first observed in~\cite{Edw99A}. Our result thus supports 
both the polarization--condensation equivalence conjecture of Ref.~\cite{Ale12A}, 
and the possibility that $T_{ch}>T_c$ in quenched QCD.

\section{Discussion}
``Bottom--up'' approach to QCD vacuum structure starts with identifying the features 
associated with important vacuum effects, such as confinement or SChSB, and 
characterizing them in a model--independent manner. Here we focused on SChSB and found
that it is intimately tied to {\em dynamical local chirality} of low--lying Dirac 
modes. In particular, strong interaction supports the band of chirally polarized 
low--energy modes that condense, and are thus ``carriers'' of the breaking phenomenon. 
The width of the band $\Lambda_{ch}$ provides for a new dynamical scale associated with 
SChSB. Simultaneous occurrence of mode condensation and chiral polarization at finite 
temperature supports the possibility that chiral symmetry breakdown 
(true and ``valence'') is equivalent to presence of chiral polarization,
and thus non--zero $\Lambda_{ch}$~\cite{Ale12A}. This would extend the equivalence
of Eq.~(\ref{eq:10}) to include the independent condition 
$\lim_{m\to 0}\Lambda_{ch}(m)>0$.

\section*{References}
\bibliographystyle{iopart-num}
\bibliography{pap1}

\end{document}